\documentclass{article}

\usepackage{microtype}
\usepackage{graphicx}
\usepackage{subfigure}
\usepackage{booktabs} 

\usepackage{hyperref}

\usepackage[accepted]{icml2024}

\usepackage{amsmath}
\usepackage{amssymb}
\usepackage{mathtools}
\usepackage{amsthm}

\usepackage[capitalize,noabbrev]{cleveref}

\theoremstyle{plain}

\theoremstyle{definition}

\theoremstyle{remark}

\usepackage[textsize=tiny]{todonotes}

\icmltitlerunning{EggNet Particle Track Reconstruction}

\begin{document}

\twocolumn[
\icmltitle{EggNet: An Evolving Graph-based Graph Attention Network for\\ Particle Track Reconstruction}



\icmlsetsymbol{equal}{*}

\begin{icmlauthorlist}
\icmlauthor{Paolo Calafiura}{lbnl}
\icmlauthor{Jay Chan}{lbnl}
\icmlauthor{Loic Delabrouille}{fr,lbnl}
\icmlauthor{Brandon Wang}{ucb,lbnl}
\end{icmlauthorlist}

\icmlaffiliation{lbnl}{Scientific Data Division, Lawrence Berkeley National Laboratory, Berkeley, CA 94720, USA}
\icmlaffiliation{fr}{Department of Computer Science, École normale supérieure de Rennes, 35170 Bruz, France}
\icmlaffiliation{ucb}{Department of Computer Science, University of California, Berkeley, CA 94720, USA}

\icmlcorrespondingauthor{Jay Chan}{jaychan@lbl.gov}

\icmlkeywords{Machine Learning, ICML}

\vskip 0.3in
]



\printAffiliationsAndNotice{\icmlEqualContribution} 

\begin{abstract}
Track reconstruction is a crucial task in particle experiments and is traditionally very computationally expensive due to its combinatorial nature. Recently, graph neural networks (GNNs) have emerged as a promising approach that can improve scalability. Most of these GNN-based methods, including the edge classification (EC) and the object condensation (OC) approach, require an input graph that needs to be constructed beforehand. In this work, we consider a one-shot OC approach that reconstructs particle tracks directly from a set of hits (point cloud) by recursively applying graph attention networks with an evolving graph structure. This approach iteratively updates the graphs and can better facilitate the message passing across each graph. Preliminary studies on the TrackML dataset show better track performance compared to the methods that require a fixed input graph.
\end{abstract}

\section{Introduction}
Track reconstruction is an essential task in particle experiments. For each particle collision event\footnote{A particle collision event is typically a bunch crossing of two particle beams with opposite directions, resulting in numerous particles emerging from the collision point.}, the goal of track reconstruction is to associate a list of 2D or 3D position measurements from a tracking detectors (referred to as ``hits'') to a list of particle track candidates. A track candidate is formed by a list of hits that are determined by the tracking algorithm. Ideally, the hits in each track candidate should originate from the same particle, and the tracking algorithm should identify all of the particles emerging from the collisions (i.e. 100\% efficiency).

The traditional tracking algorithm, the Combinatorial Kalman Filter \cite{RevModPhys.82.1419, ATLAS:2017kyn, CMS:2014pgm}, starts with a track seed of two (doublet) or three (triplet) hits, each of which provides an initial track direction. The next possible hits are then added by iteratively looking for the hits that match the extrapolated trajectory. Despite robust track performance, this method is computationally expensive and the computational cost does not scale well with the number of hits (referred to as ``event size''). Performing such tracking algorithm thus becomes much more challenging in the future experiments (such as HL-LHC), where the event size is significantly increased.

In order to address the computational challenge of the traditional algorithm, recent developments have focused on applying graph neural networks (GNNs) to the tracking procedure \cite{ExaTrkX:2021abe, Biscarat:2021dlj, CHEP2022, Lieret:2023aqg} and demonstrated that the GNN-based tracking algorithms can achieve linear scaling with the event size \cite{ExaTrkX:2021abe, Lazar:2022ixi}. These GNN-based tracking algorithms can be classified into two approaches: edge classification (EC)  \cite{ExaTrkX:2021abe, Biscarat:2021dlj, CHEP2022} and object condensation (OC) \cite{Lieret:2023aqg}. In the EC approach, the GNN is trained to remove edges (based on the learned edge scores) that connect hits belonging to different particles. This will then be followed by a graph segmentation algorithm, which assigns each connected component as a track candidate. In the OC approach, the goal of the GNN is to learn the node representation in a latent space, where hits belonging to the same particles are attracted to one another. One can then perform a clustering for the hits in the learned latent space and assign each cluster as a track candidate.

While effective, most of these methods require a graph construction step before performing GNN. Since it is unrealistic to construct a fully connected graph ($N_\mathrm{edges} \sim N_\mathrm{nodes}^2$ ), multiple methods, such as metric learning \cite{ExaTrkX:2021abe} and module map \cite{Biscarat:2021dlj} have been developed to construct a graph connecting the hits that are likely to belong to the same particles. In metric learning, embedding for each node is learned by a multi-layer perception (MLP). Edges are then constructed, connecting the nearest neighbors in the embedding space. Here, the embedding for each node is purely based on the features of that node and does not depend on how the node interacts with others. Due to the lack of awareness of other nodes, the metric-learning-based graph construction tends to produce a graph that contains many fake edges (low edge purity). One often needs to use another network that takes edge features as inputs and further reduces the graph size by removing edges \cite{ExaTrkX:2021abe}. In the module map approach, one constructs the graph based on all possible connections between detector modules. This is currently resource-intensive and also tends to produce low-purity graphs.

It is important to note that graph construction efficiency plays a crucial role in the performance of later stages for both EC and OC approaches. In the EC approach, any graph construction inefficiency (the missing true edges in the constructed graph) is directly propagated to the graph segmentation stage. In the OC approach, it is possible to recover missing true edges after the clustering. However, graph construction inefficiency still leads to information loss during the GNN message passing step. Furthermore, low purity (due to fake edges) results in noisy information during the message passing. Both cases affect the effectiveness of the message passing step.

In this work, we propose an OC-based method that embeds the graph construction into an Evolving Graph-based Graph Attention Network (EggNet) architecture, which takes point clouds as inputs and iteratively constructs the graphs based on the updated embedding. The message passing in the EggNet is then based on the updated graph in each iteration. This method allows to gradually enhance the constructed graph efficiency in each iteration and thus can improve the effectiveness of the message passing. We test the method with the TrackML dataset\footnote{TrackML dataset is a dataset commonly used for benchmarking the performance of a tracking algorithm. It contains collision events simulated with a general particle detector. The event size in the TrackML dataset is comparable to HL-LHC.} \cite{Amrouche:2019wmx, Amrouche:2021nbs} and obtain better track performance compared to the existing GNN-based methods that require a pre-constructed input graph.


\section{Method}
\begin{figure*}[ht]
\vskip 0.2in
\begin{center}
\centerline{\includegraphics[width=\textwidth]{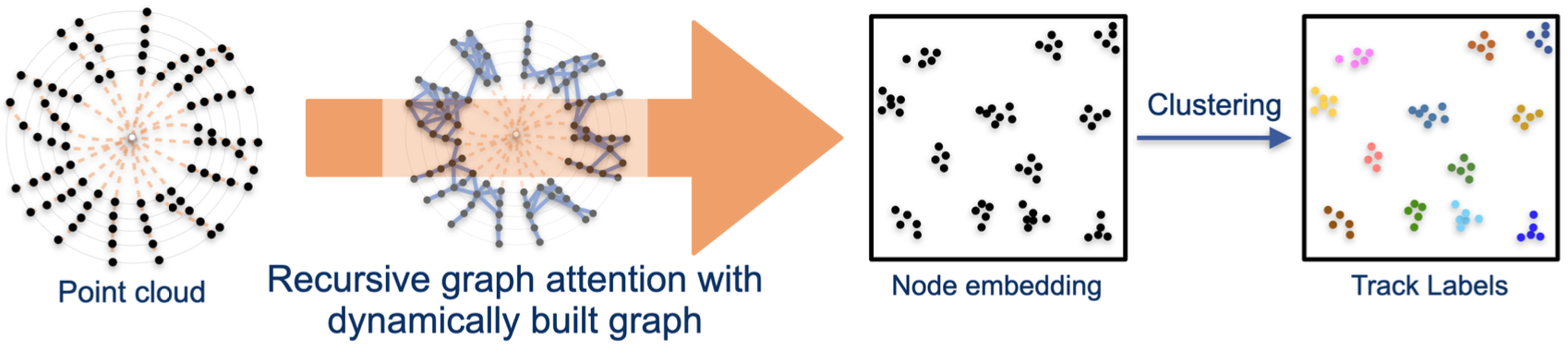}}
\caption{The proposed end-to-end tracking pipeline. Starting with a point cloud, where each point corresponds to a hit, we run an EggNet model that iteratively constructs a graph and performs message passing. EggNet outputs the embedding for each node, and the track candidates are extracted from the DBSCAN clusters that are obtained in the node embedding space.}
\label{fig:pipeline}
\end{center}
\vskip -0.2in
\end{figure*}

The proposed end-to-end tracking pipeline consists of an EggNet step and a clustering step. As shown in \cref{fig:pipeline}, the EggNet takes a set of hits (a point cloud) as inputs and outputs the node embedding for each node. We adopt a similar approach to GravNet \cite{Qasim:2019otl}, where the graph attention acts on dynamically built KNN graphs. In comparison with \cite{Qasim:2019otl, veličković2018graph}, our attention weight is learned from a dedicated edge network, and the aggregation is based on the learned edge representation rather than the neighboring node representation. The nodes are then clustered in the node embedding space and assigned to different track candidates. In our study, we adopt Density-Based Spatial Clustering of Applications with Noise (DBSCAN) \cite{10.5555/3001460.3001507} as the clustering algorithm.

\subsection{EggNet}

\begin{figure*}[htbp]
\vskip 0.2in
\begin{center}
\includegraphics[width=0.54\textwidth]{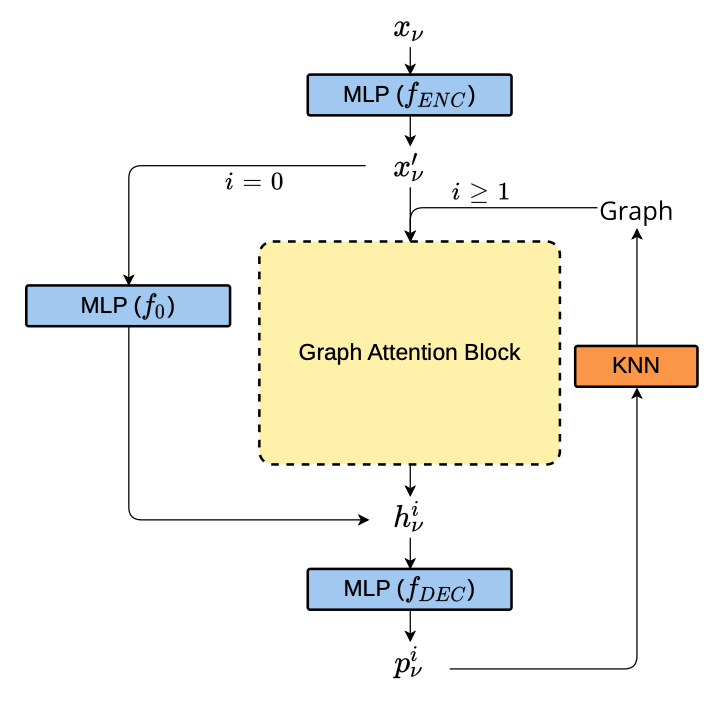}
\includegraphics[width=0.44\textwidth]{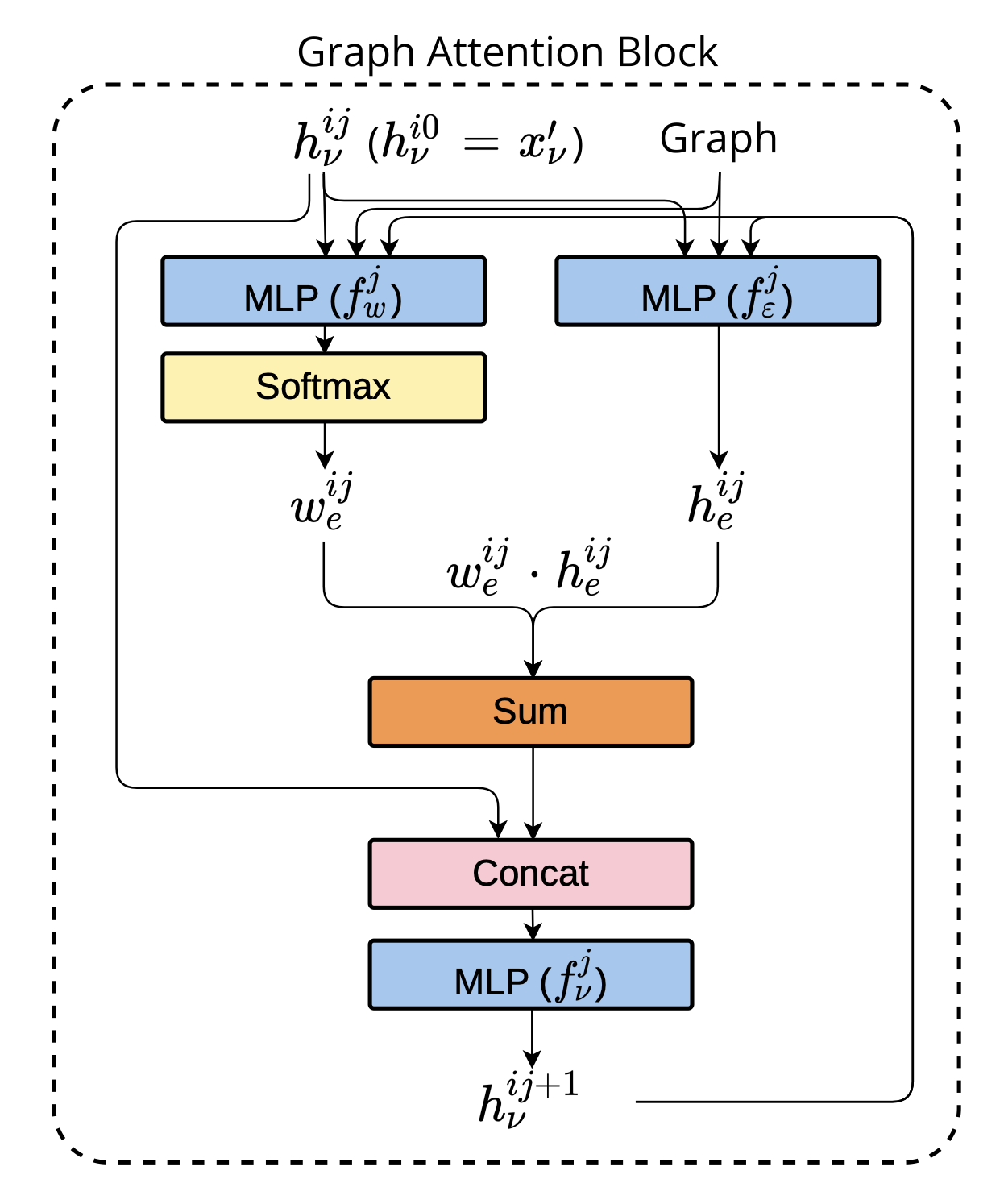}
\caption{The EggNet architecture. $i$ corresponds to each EggNet iteration, and $j$ corresponds to each message passing step. An EggNet iteration generally consists of a graph attention block and a KNN. The first iteration does not perform graph attention and the last iteration does not perform KNN.}
\label{fig:RGAT}
\end{center}
\vskip -0.2in
\end{figure*}

The EggNet architecture is illustrated in \cref{fig:RGAT}. The $d$-dimensional input node features $\nu \in \mathbb{R}^d$ are first encoded:
\begin{equation}
x \in \mathbb{R}^{d_x} = f_\mathrm{ENC} \left(\nu \right),
\end{equation}
where $f_\mathrm{ENC}$ is the MLP node encoder. The encoded node features are then taken as inputs by a number of iterations where we first update the graph node embeddings and then the edges by connecting the K-Nearest Neighbors (KNN) in the node embedding space. In the first iteration ($i=0$), each node is projected into a latent space (denoted as $h_0$-space) by an MLP node network:
\begin{equation}
h_0 \in \mathbb{R}^{d_h} = f_0 \left(x \right).
\end{equation}
A MLP node decoder $f_\mathrm{DEC}$ then acts on $h_0$ and outputs a node embedding in $p_0$-space:
\begin{equation}
p_0 \in \mathbb{R}^{d_p} = f_\mathrm{DEC} \left(h_0 \right).
\end{equation}
This is then followed by a KNN in $p_0$ which constructs the first graph $G_0$.

In each of the next iterations ($i \geq 1$), multiple graph-attention-based \cite{veličković2018graph} message passing steps are performed to obtain the updated node representation $h_i \in \mathbb{R}^{d_h}$. The same node decoder $f_\mathrm{DEC}$ is then used to obtain the updated node embedding in $p_i$-space:
\begin{equation}
p_i \in \mathbb{R}^{d_p} = f_\mathrm{DEC} \left(h_i \right),\ i \geq 1,
\label{eq:decoder}
\end{equation}
which is followed by a KNN that updates the graph $p_i \to G_i$.

In each message passing step, we first learn an edge representation ($e_{ij} \in \mathbb{R}^{d_e}$, where $i$ denotes the current EggNet iteration and $j$ denotes the current message passing step), and an edge weight $w_{ij} \in \mathbb{R}$ from the two connecting nodes for each edge:
\begin{equation}
e_{ij} =
\begin{cases}
f_\epsilon^j\left(x^\mathrm{S}, x^\mathrm{T} \right),& i \geq 1,\ j = 0\\
f_\epsilon^j\left(h_{ij}^\mathrm{S}, h_{ij}^\mathrm{T}, e_{i(j-1)} \right),& i \geq 1,\ j \geq 1
\end{cases},
\end{equation}
and
\begin{equation}
w_{ij} =
\begin{cases}
f_w^j\left(x^\mathrm{S}, x^\mathrm{T} \right),& i \geq 1,\ j = 0\\
f_w^j\left(h_{ij}^\mathrm{S}, h_{ij}^\mathrm{T}, e_{i(j-1)} \right),& i \geq 1,\ j \geq 1
\end{cases},
\end{equation}
where S and T denote the source and target nodes of an edge, respectively. $h_{ij}$ is the current representation of a given node. Note that for $j \geq 1$, in addition to the two connecting nodes, the edge networks $f_\epsilon^j$ and $f_w^j$ also take the previous edge representation $e_{i(j-1)}$ as inputs.

The node representation $h_{ij}$ is updated by a node network $f_\nu^j$ in each message passing step, which takes as the inputs the weighting sum of all connecting edges for each node:
\begin{equation}
    a_{ij} = \sum_{\epsilon \in E} \text{softmax} \left( w_{ij}^\epsilon \right) e_{ij}^\epsilon,
\end{equation}
\begin{equation}
    h_{ij} =
\begin{cases}
f_\nu^j \left(x,\ a_{i0} \right),& i \geq 1,\ j = 0\\
f_\nu^j \left(h_{i(j-1)},\ a_{ij} \right),& i \geq 1,\ j \geq 1
\end{cases},
\end{equation}
where $E$ corresponds to all connecting edges for a given node. The $h_{ij}$ from the last message passing step ($h_{i(-1)}$) is then taken as $h_i$ that enters \cref{eq:decoder}.

\subsection{Loss Function}
The EggNet model is optimized to minimize the loss $L$:
\begin{equation}
    L = \mathbb{E}_{\epsilon \in \text{true\ edges}}\left(l_\epsilon \right) + \mathbb{E}_{\epsilon \in \text{random\ edges}} \left(l_\epsilon \right) + \mathbb{E}_{\epsilon \in \text{KNN\ edges}} \left(l_\epsilon \right),
\label{eq:loss}
\end{equation}
where $l_\epsilon$ is the individual loss term for a given representative edge connection:
\begin{equation}
    l_\epsilon = y_\epsilon d_\epsilon^2 + \left(1-y_\epsilon\right){\max}^2\left(0, m-d_\epsilon\right),
\end{equation}
where the first term is an attractive loss that brings together nodes belonging to the same particles, and the latter is a repulsive hinge loss that separates nodes belonging to different particles. $y_\epsilon$ is the truth label for each edge, which is 1 for an edge connecting hits coming from the same particle (true edge), and 0 otherwise (fake edge). $d_\epsilon$ is the Euclidean distance between two nodes in the $p_j$-space in the last EggNet iteration ($p_{-1}$), and $m$ is a constant value, which is set to 1 in our study. The representative edges are pooled from all true edges, randomly selected edges and all KNN edges, corresponding to the three terms in \cref{eq:loss}. Since the majority of randomly selected edges are fake edges, the combination of true and randomly selected edges ensure that we optimize the model based on both true and fake edges. KNN edges represent the most true-like edges (edges connecting the nodes that are the nearest). Sampling from these edges allow us to emphasize the fake edges that look like true edges.

\subsection{DBSCAN}
$p_{-1}$ is taken as inputs of DBSCAN, which then outputs a track label $\lambda$ for each node:
\begin{equation}
    \{\lambda_\nu \} = \mathrm{DBSCAN}\left(\{p_{-1} \} \right) | \nu \in V,
\end{equation}
where V represents all nodes in a graph, i.e., all hits in a collision event. All meaningful track labels are positive integers, and all nodes that share the same meaningful track label form a track candidate. For DBSCAN, it is possible that a node does not belong to any cluster. In this case, the $\lambda$ is assigned as 0 and the node is considered as a noise hit.

\subsection{Models and Implementation}

All MLPs in the EggNet model consist of 3 hidden layers, each with a width of 128 neurons. Each intermediate layer in these networks uses a SiLU \cite{ELFWING20183} activation function and layer normalization \cite{ba2016layer}. The last layer of $f_\mathrm{DEC}$ uses a Tanh activation function, and the output is further normalized $p^\prime = \frac{p}{|p|}$. For the following studies, we set $d_x$, $d_h$ and $d_e$ to 128, while $d_p$ is set to 24. The EggNet model consists of up to 5 EggNet iterations ($i \leq 4)$, and each graph attention block consists of 8 message passing steps ($j \leq 7$). We consider $k=10$ for all KNN steps in the training.

All neural networks are implemented using PyTorch \cite{NEURIPS2019_9015} and PyTorch Lightning \cite{Falcon_PyTorch_Lightning_2019} in the ACORN framework \cite{acorn} and simultaneously optimized with Adam \cite{adam} with a learning rate of $2 \times 10^{-4}$. The training uses a batch size of 1 event and is performed for 200 epochs.

DBSCAN is implemented using Rapids cuML \cite{rapids}.

\section{Experimental Setup and Evaluation}

In the following, we test our tracking pipeline with the TrackML dataset, which is a simulation of a generic tracking detector under the HL-LHC pileup condition. For simplicity, we consider only particles with transverse momentum $p_\mathrm{T} > 1$ GeV. This means hits coming from particles with $p_\mathrm{T} < 1$ GeV and noise hits are removed from our simulated datasets, reducing the number of hits by a factor of $\sim 10$. Note that applying such a selection on the transverse momentum significantly reduces the event size, making the task of track reconstruction much easier. It is therefore important to study the performance with the full events (no hits are removed). We leave the study with full events for future work.

We train the EggNet model with 12 input node features $\nu \in \mathbb{R}^{12}$ (described in \cite{ExaTrkX:2021abe}). We first evaluate the EggNet edge-wise performance on the KNN graphs built in the $p_{-1}$-space. We define two metrics for the KNN edge-wise performance, including ``edge-wise efficiency'' $\mathrm{Eff_{KNN}}$ and ``edge-wise purity'' \footnote{``Efficiency'' and ``purity'' are also commonly known as ``recall'' and ``precision''.} $\mathrm{Pur_{KNN}}$:

\begin{equation}
    \mathrm{Eff_{KNN}} = \frac{N_\mathrm{KNN}^{y=1}}{N^{y=1}},
\end{equation}

\begin{equation}
    \mathrm{Pur_{KNN}} = \frac{N_\mathrm{KNN}^{y=1}}{N_\mathrm{KNN}} = \frac{N_\mathrm{KNN}^{y=1}}{N_\mathrm{hits} \cdot k},
\end{equation}

where $N^{y=1}$ is the number of all true edges, $N_\mathrm{KNN}$ is the number of edges counted from all KNN graphs, which is equal to $N_\mathrm{hits} \cdot k$, and $N_\mathrm{KNN}^{y=1}$ is the number of true KNN edges. The EggNet performance is the best when $\mathrm{Eff_{KNN}}$ and $\mathrm{Pur_{KNN}}$ are closest to 1 (they cannot be larger than 1). Note that $N_\mathrm{KNN}^{y=1}$ cannot exceed $k$ per node or the number of true edges for each source node:

\begin{equation}
    N_\mathrm{KNN}^{y=1} \leq \sum_{\nu^S} \min \left(k,\ N^{y=1}_{\nu^S} \right).
\end{equation}

\begin{figure}[htbp]
\begin{center}
\includegraphics[width=\columnwidth]{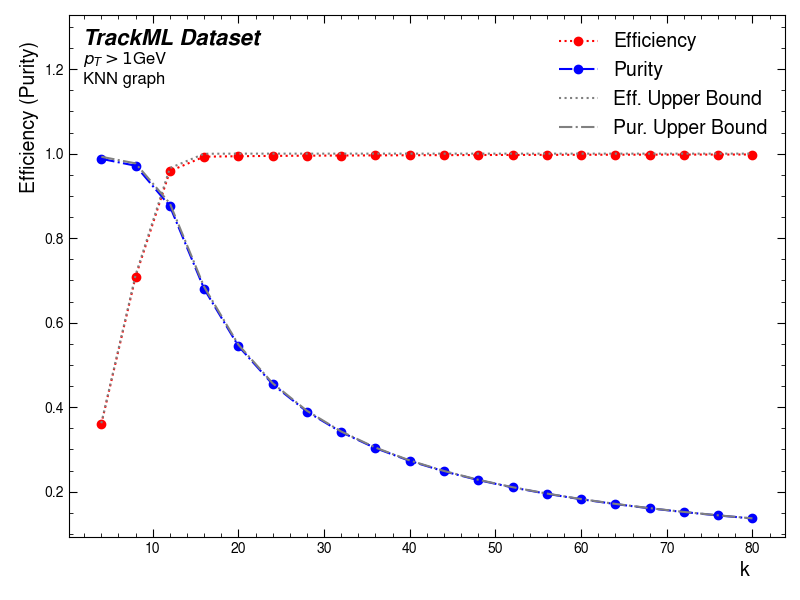}
\caption{Edge-wise efficiency and purity in the KNN graphs as a function of $k$. The EggNet here consists of 5 iterations ($i \leq 4$). The upper bound for each metric is also shown in the plot.}
\label{fig:edge_wise_eff_vs_k}
\end{center}
\end{figure}

This thus determines the upper bounds of $\mathrm{Eff_{KNN}}$ and $\mathrm{Pur_{KNN}}$. As shown in \cref{fig:edge_wise_eff_vs_k}, we plot $\mathrm{Eff_{KNN}}$ and $\mathrm{Pur_{KNN}}$ as a function of $k$. $\mathrm{Eff_{KNN}}$ increases and approaches 1 as $k$ increases, while $\mathrm{Pur_{KNN}}$ decreases as $k$ increases. Upper bounds of $\mathrm{Eff_{KNN}}$ and $\mathrm{Pur_{KNN}}$ are also shown in the figure. We observe that both $\mathrm{Eff_{KNN}}$ and $\mathrm{Pur_{KNN}}$ are very close to the upper bounds for any $k$, which indicates that the EggNet performance is close to optimal. Note that the KNN graphs are only used as the inputs for the message passing and training loss, and they do not correspond to the final track candidates which are obtained using DBSCAN. Thus, the performance on the KNN graphs does not directly translate to the track performance. However, it allows us to evaluate the EggNet model performance with effects decoupled from the later stage of DBSCAN.

\begin{figure}[htbp]
\begin{center}
\includegraphics[width=\columnwidth]{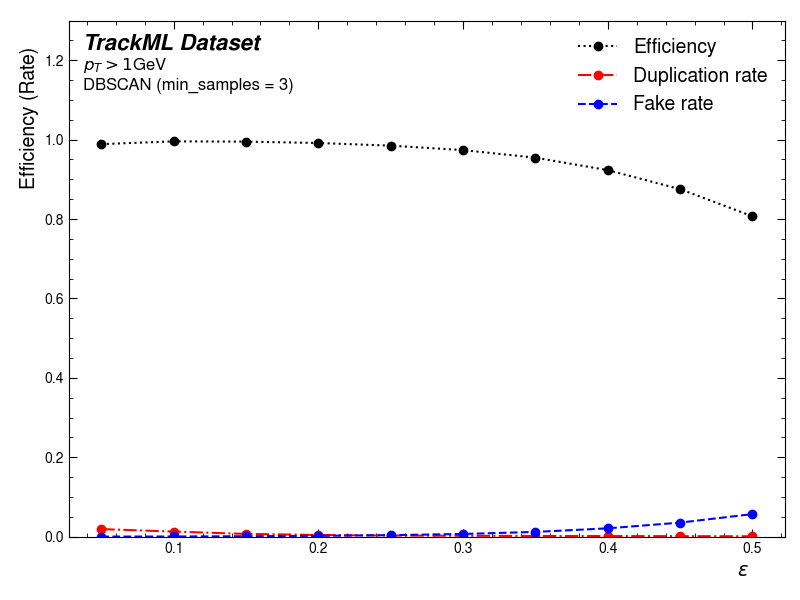}
\caption{Track efficiency, duplication rate and fake rate with track candidates obtained from the DBSCAN clusters as a function of $\epsilon$. The EggNet here consists of 5 iterations ($i \leq 4$).}
\label{fig:dbscan_vs_epsilon}
\end{center}
\end{figure}

The track performance is then evaluated on all track candidates extracted from the DBSCAN clusters. A track candidate is matched to a particle if more than 50\% of the track hits come from that particle. Three metrics are then defined as follows:
\begin{itemize}
    \item Efficiency: The fraction of particles to which at least 1 track candidate is matched.
    \item Duplication rate: The fraction of track candidates that are matched to the same particles.
    \item Fake rate: The fraction of track candidates that are not matched to any particle.
\end{itemize}
\cref{fig:dbscan_vs_epsilon} shows the three metrics as a function of $\epsilon$, which is a parameter of DBSCAN. In particular, when $\epsilon = 0.1$, we obtain a track efficiency of 0.9956, duplication rate of 0.0129 and fake rate of 0.0006. Note that the choice of $\epsilon = 0.1$ here is made to maximize the track efficiency with the trade off of slightly higher duplication and fake rates.

For comparison, we also perform track reconstruction with both the EC and OC approaches where a pre-constructed graph is required as an input for the graph attention network. For these pre-constructed graph-based approaches, we adopt the metric learning method \cite{ExaTrkX:2021abe} for graph construction. We simply take the same architecture as in the first EggNet iteratioin ($i=0$), which is an MLP, as the neural network used in metric learning and perform KNN with $k=10$. The graph attention network has the same architecture as the graph attention block in the EggNet model for both EC and OC approaches. For each scenario, we estimate the uncertainty by training 5 independent models and taking the standard deviation. The median is taken as the nominal result. A comparison with different number of EggNet iterations, and with methods that require a pre-constructed input graph, is shown in \cref{tab:comp}. We observe that EggNet starts outperforming the pre-constructed graph-based methods when the number of EggNet iterations is greater than 3 ($i \leq 2$).

\begin{table*}[t]
\caption{Track performance with EggNet with various number of EggNet iterations ($i$). $\epsilon = 0.1$ is used for the DBSCAN in all scenarios. The performance is compared with methods that require pre-constructed input graphs, including the edge classification approach (EC) and the object condensation approach (OC). The uncertainty in each scenario is estimated with the standard deviation of 5 independent models, and the median is taken as the nominal result.}
\label{tab:comp}
\vskip 0.15in
\begin{center}
\begin{small}
\begin{sc}
\begin{tabular}{lcccr}
\toprule
Method & Efficiency & Dup. Rate & Fake Rate \\
\midrule
EC    & $0.9898 \pm 0.0009$ & $0.0421 \pm 0.0011$ & $0.0012 \pm 0.0000$ \\
OC & $0.9902 \pm 0.0007$ & $0.0328 \pm 0.0007$ & $0.0015 \pm 0.0001$ \\
EggNet ($i \leq 1$)    & $0.7454 \pm 0.0027$ & $0.2202 \pm 0.0024$ & $0.0099 \pm 0.0004$ \\
EggNet ($i \leq 2$)    & $0.9905 \pm 0.0005$ & $0.0179 \pm 0.0006$ & $0.0011 \pm 0.0000$ \\
EggNet ($i \leq 3$)     & $0.9940 \pm 0.0002$ & $\mathbf{0.0117 \pm 0.0003}$ & $\mathbf{0.0005 \pm 0.0000}$ \\
EggNet ($\mathbf{i \leq 4}$)      & $\mathbf{0.9956 \pm 0.0003}$ & $0.0129 \pm 0.0002$ & $0.0006 \pm 0.0000$ \\
\bottomrule
\end{tabular}
\end{sc}
\end{small}
\end{center}
\vskip -0.1in
\end{table*}

\begin{figure}[htbp]
\begin{center}
\includegraphics[width=\columnwidth]{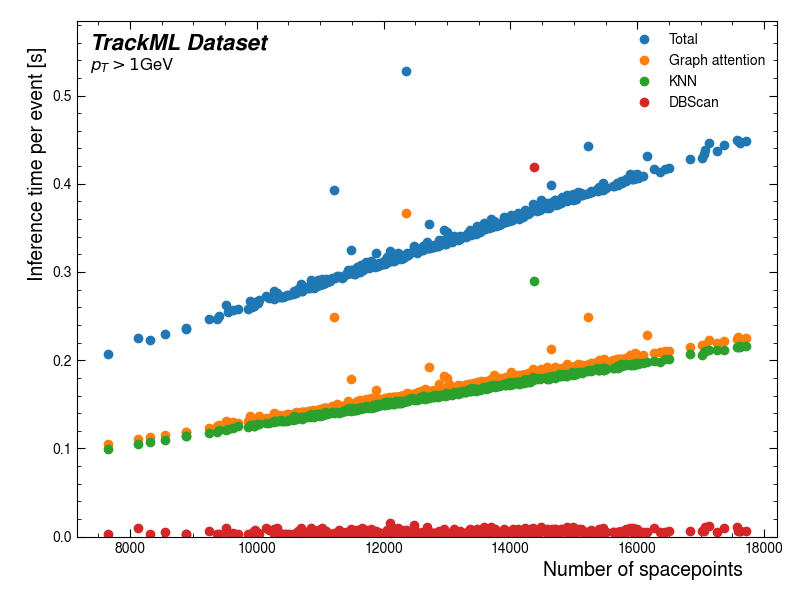}
\caption{Inference time for each event versus the number of hits. The inference time is also broken down different components, including Graph Attention, KNN and DBSCAN. The EggNet here consists of 5 iterations ($i \leq 4$).}
\label{fig:computing_time}
\end{center}
\end{figure}

Finally, \cref{fig:computing_time} shows the inference time when running on an NVIDIA A100  Graphical Processing Unit (GPU) versus the total number of hits for each event. The averaged total computing time per event is around 0.26 seconds. The total computing time is also broken down into different components, including Graph Attention, KNN and DBSCAN. The majority of the computing time comes from graph attention and KNN.

\section{Conclusions}

In this paper, we proposed a new object condensation-based approach to particle track reconstruction, termed EggNet. EggNet takes point clouds as inputs and builds graphs on the fly for each EggNet iteration. The constructed graphs are iteratively updated, improving message-passing efficacy. We demonstrated this approach with the TrackML dataset and obtained excellent track performance in both KNN graphs and reconstructed track candidates. EggNet also outperforms the pre-constructed graph-based tracking methods when sufficient EggNet iterations are performed. The result is encouraging and motivates future work to improve the computational performance. In particular, KNN has a significant contribution to the computing time. It will be beneficial to reduce the time consumption with an alternative method, such as approximate nearest neighbors and radius nearest neighbors. Furthermore, the performance with the full events where no hits are removed needs to be studied, and, if necessary, improved.




\section*{Impact Statement}

The development of machine learning-based and in particular GNN-based particle track reconstruction algorithms provides significant potential for positive broader impacts in both the field of particle physics and society at large. By utilizing GNN as part of the track reconstruction pipeline, it is possible for the computational cost to scale linearly with the event size, which accelerates the discovery potential of future particle experiments such as HL-LHC. The method's potential benefits include enhancing our understanding of fundamental particles and their interactions, thereby advancing scientific knowledge and contributing to the development of more precise experimental techniques in particle physics research. Furthermore, EggNet has the potential to extend beyond particle physics, with applications in various domains that involve pattern recognition and representation learning. For instance, it could find applications in medical imaging, where accurate reconstruction of complex image structures from noisy data is crucial.


\bibliography{mybib}
\bibliographystyle{icml2024}




\end{document}